\documentstyle[aps,prb,epsf,multicol]{revtex}
\draft

\newcommand{\cntr}[1]{{\vspace*{\fill}\hspace*{\fill} #1
  \hspace*{\fill}\vspace*{\fill}}}

\begin{document}

\author{Xiao-jun Li and M. Schick\\
        Department of Physics, Box 351560 \\
        University of Washington, Seattle 98195-1560}
\title{Distribution of lipids in non-lamellar phases of their mixtures}

\date{\today}

\maketitle

\begin{abstract}
We consider a model of lipids in which a head group, characterized by its
volume, is attached to two flexible tails of equal length.  The phase
diagram of the anhydrous lipid is obtained within self-consistent field
theory, and displays, as a function of lipid architecture, a progression
of phases: body-centered cubic, hexagonal, gyroid, and lamellar.  We
then examine mixtures of an inverted hexagonal forming lipid and a
lamellar forming lipid.  As the volume fractions of the two lipids vary,
we find that inverted hexagonal, gyroid, or lamellar phases are formed.
We demonstrate that the non-lamellar forming lipid is found
preferentially at locations which are difficult for the lipid tails to
reach. Variations in the volume fraction of each type of lipid tail are
on the order of one to ten per cent within regions dominated by the
tails.  We also show that the variation in volume fraction is correlated
qualitatively with the variation in mean curvature of the head-tail
interface.
\end{abstract}
\section{Introduction}
The lipid bilayer which provides the basic structure of biological
membranes is composed of a large number of different lipids, of which
many, on their own, form non-lamellar phases.  Just what role these 
non-lamellar forming lipids play in the properties of membranes has been the
subject of much speculation\cite{cullis,dekruijff,epand}. That they
serve important functions is indicated by the fact that cells with
common lipids regulate their composition to maintain a proper balance
between those that form lamellar phases and those that do not\cite{morein}.
At least two major roles for non-lamellar forming lipids have been
proposed. One is that,
because such lipids are characterized by tails which tend to
splay outward, their presence alters the pressure profile within a
bilayer permitting embedded proteins to function\cite{marsh}. The
second, noting that the lipid balance referred to above is close to a
lamellar non-lamellar phase transition \cite{morein,rietveld},
posits that these lipids with their splaying tails 
may serve to facilitate the formation of structures
not unlike the non-lamellar phases, structures 
which are characterized by volumes
which are difficult for other lipid tails to fill, or 
regions of non-zero curvature. In this way they could stabilize
transient fusion intermediates. For example, some
scenarios of membrane fusion involve the formation of a thin cylindrical
stalk\cite{markin,siegel1,chern1} created from the lipids of the 
opposing 
leaves of the bilayers. Such a structure has a non-zero curvature, and 
creates regions surrounding it which are difficult for the tails to
fill\cite{siegel1}. By concentrating in such regions, 
non-lamellar forming lipids could
make such structures, and the processes they bring about, 
less costly\cite{siegel1,chern}. In a series of experiments, Gruner and
co-workers have shown that inverted hexagonal, $H_{II}$, phases which
are characterized by periodic regions which are difficult to fill can be
stabilized by the addition of alkanes\cite{kirk}. Presumably
the alkanes are found preferentially in these regions and lower the free
energy to form these structures. Theoretical confirmation of this idea in 
the somewhat analogous system of block
copolymer and homopolymer mixtures 
was provided by Matsen\cite{matsenmix}. Tate and
Gruner\cite{tate} further showed that the addition of small amounts of
long chain (two tails of 22 or 24 carbons) phosphatidylcholine to 
dioleoylphosphatidylethanolamine (two tails of 18 carbons) stabilized the
$H_{II}$ phase relative to that produced by the addition of 
phosphatidylcholine with two shorter tails of 18 carbons.
This provided further indirect evidence for the common hypothesis
that
difficult to fill, ``frustrated" volumes and/or regions of high
curvature\cite{andelman} should be correlated with a density difference between
lamellar- and non lamellar-forming lipids. However  there appears to be neither
experimental evidence nor theoretical calculations that directly bear on
this hypothesis, or provide an indication of the magnitude of the
variation in density of the different lipids.
It is the purpose of this paper to demonstrate and to quantify, in a
model lipid mixture, this variation of density of the lamellar- and non
lamellar-forming lipids in inverted hexagonal, $H_{II}$, and gyroid
($Ia3d$) phases.

Our paper is organized as follows. In section II, we present the model
of the lipids and develop the self consistent field theory in real
space. In section III we develop the self consistent field theory in
Fourier space. In section IV(A) we present our results for the phase
diagrams of the single lipid as a function of its architecture, and for
the anhydrous mixture of a lamellar forming and a non-lamellar forming
lipid. We also show that there exists an ``effective single lipid
approximation" by which the latter phase diagram can be obtained from
the former. In section IV(B) we present our results for the variation of
the densities of the two lipids in the lamellar, inverted hexagonal, and
gyroid phases. This variation is on the order of 1 to 10\%. We also
compare qualitatively this variation with that of the mean curvature of
the structures.
\section{Theory: Real Space}
The model which we employ has been presented elsewhere\cite{li}, so we
will be brief here. We consider an anhydrous mixture of $n_1$ lipids of
type 1 and $n_2$ lipids of type 2. Below we shall choose their
architecture so that type 1 lipids form lamellar phases while type 2
lipids form $H_{II}$ phases. All lipids consist of the {\it same} head group
of volume $v_h$ and two equal-length tails. Thus we model 
mixtures of  lipids drawn from a homologous series, such as the
phosphatidylethanolamines studied by Seddon et al.\cite{scm}.  Each
tail of lipid 1 consists of $N_1$ segments 
of volume $v_t$, while those of
lipid 2 consist of $N_2$ such segments. 
For convenience, we denote $N_L=N\alpha_L$ for $L=1,2$. 
The tails are treated as
being completely flexible, with radii of gyration
$R_{g,L}=(N_La^2/6)^{1/2}$ for each tail.
The statistical segment length is $a$. 
The configuration of the 
$l$'th lipid of type $L$ is described by a space curve ${\bf
r}_{l,L}(s)$ where $s$ ranges from 0 at one end of one tail, through
$s=\alpha_L/2$ at which the head is located, to $s=\alpha_L$, 
the end of the other tail.
The system is completely
described by the dimensionless 
densities of the head groups $\hat\Phi_h^{(L)}({\bf
r})$ and of the tail segments, $\hat\Phi_t^{(L)}({\bf r})$, which can be
written
\begin{eqnarray}
\hat\Phi_h^{(L)}({\bf r})&=&v_h\sum_{l=1}^{n_L}\delta\left({\bf r}-
{\bf r}_{l,L}(\alpha_L/2)\right),\\
\hat\Phi_t^{(L)}({\bf r})&=&v_h\sum_{l=1}^{n_L}\int_0^{\alpha_L}
\delta\left({\bf r}-
{\bf r}_{l,L}(s)\right)ds,
\end{eqnarray}
where $v_h$ has been chosen as a convenient volume to make all
densities dimensionless.
The number density of tail segments of type $L$ is 
$(2N/v_h)\hat\Phi_t^{(L)}$, and their volume fraction is 
$(2Nv_t/v_h)\Phi_t^{(L)}\equiv\gamma_t\Phi_t^{(L)}$.
The sole explicit interaction in the system is between head
and tail segments, and this interaction energy $E$ takes the form
\begin{equation}
\label{energy}
 \frac{1}{kT}E[\hat\Phi_h^{(1)}+\hat\Phi_h^{(2)},\hat\Phi_t^{(1)}+
\hat\Phi_t^{(2)}]
\equiv\frac{2N\chi}{v_h}\int \left[\hat\Phi_h^{(1)}({\bf r)}+
\hat\Phi_h^{(2)}({\bf r})\right]\left[ \hat\Phi_t^{(1)}({\bf r})+
\hat\Phi_t^{(2)}({\bf r)}\right]d{\bf r},
\end{equation}
where 
$\chi$ is the strength of the interaction, and $T$ is the 
temperature. The effect of a hard-core repulsion between all elements of
the system is accounted for approximately by requiring that the system
be incompressible.
The grand canonical partition function of the system is\cite{matsen}
\begin{equation}
\label{pf1}
{\cal Z}=\sum_{n_1,n_2}{z_1^{n_1}z_2^{n_2}\over n_1! n_2!}\int\prod_{l=1}^{n_1}
\tilde{\cal D}{\bf r}_{l,1}\prod_{m=1}^{n_2}\tilde{\cal D}{\bf r}_{m,2}
\exp(-E/kT)\delta({\hat\Phi}_h^{(1)}+\gamma_t{\hat\Phi}_t^{(1)}
+{\hat\Phi}_h^{(2)}+\gamma_t{\hat\Phi}_t^{(2)}-1).
\end{equation}
The delta function in the above enforces the constraint of 
incompressibility. The notation
$\int\tilde{\cal D}{\bf r}_{l,L}$ denotes a functional integral over the
possible configurations of the $l$'th lipid of type $L$ and in which, in
addition to the Boltzmann weight, the path is weighted by the factor
${\cal P}[{\bf r}_{l,L}(s);0,\alpha_L]$, with
\begin{equation}
{\cal P}[{\bf r},s_1,s_2]={\cal N}\exp\left[-{1\over
8R_g^2}\int_{s_1}^{s_2}ds|{d{\bf r}(s)\over ds}|^2\right],
\end{equation}
where ${\cal N}$ is an unimportant normalization constant and $R_g\equiv
(Na^2/6)^{1/2}$ is the radius of gyration of a tail of length $N$. 

We note that because of the choice of this weight function, ${\cal P}$,
and the lack of any explicit interaction
between chain segments to prevent their intersection, the behavior of
the chains is Gaussian. This is appropriate because we view the chains
as forming an incompressible melt.  Under such conditions, a flexible,
interacting,  polymer
chain behaves as an ideal, and therefore Gaussian, 
one\cite{edwards1,edwards2}.   
It is not clear, of course, that an ideal {\em lipid} chain, which is
certainly not flexible, can be treated as Gaussian. This approximation
must overestimate the entropy of the tails, whose fewer thermally accessible  
configurations are presumably  modelled more accurately 
by the Restricted Isomeric States
Model\cite{flory,muller}. But how serious is this overestimation, is
also unclear. Ultimately the efficacy of our model in capturing the
behavior of lipids can be judged only by a comparison
with experiment. This was done for the lipid phase behavior 
in reference\cite{li}, and the comparison was very
good. Even the variation with solvent concentration and with 
temperature of the characteristic period of the lamellar and hexagonal
phases agreed very well. It is this agreement with experiment of the
calculated phase behavior
which provides the support for applying the model to the calculation of
other properties, such as the distribution of lipids in mixtures
investigated here.  
 
To proceed, we make the partition function of Eq. \ref{pf1} more
tractable 
by introducing
 into it
the identity
\begin{eqnarray}
1&=&\int{\cal D}\Phi_h^{(L)}\delta(\Phi_h^{(L)}-\hat\Phi_h^{(L)}),\nonumber\\
 &=&\int{\cal D}\Phi_h^{(L)}{\cal D}W_h^{(L)}\exp\left\{{1\over v_h}
\int W_h^{(L)}({\bf r})[\Phi_h^{(L)}({\bf r})-\hat\Phi_h^{(L)}
({\bf r})]d{\bf r}\right\},
\end{eqnarray}
where the integration on $W_h^{(L)}$ extends up the imaginary axis. 
We also insert
such an identity for the density $\hat\Phi_t^{(L)}({\bf r})$, and a
similar representation for the delta function which enforces the
incompressibility constraint.
The 
partition function  becomes
\begin{equation}
\label{pf2} 
{\cal Z}=
\int{\cal D}\Xi\prod_{L=1}^2{\cal D}\Phi_h^{(L)}{\cal D}W_h^{(L)}{\cal D}
\Phi_t^{(L)}{\cal D}W_t^{(L)}\exp[-\Omega/kT],
\end{equation}
where the grand potential $\Omega$ is given by
\begin{eqnarray}
\Omega&=&-\frac{kT}{v_h}\sum_{L=1}^2\left\{z_L{\cal
Q}_L[W_h^{(L)},W_t^{(L)}] +\int d{\bf r}\left[W_h^{(L)}({\bf
r})\Phi_h^{(L)}({\bf r})+W_t^{(L)}({\bf r})
\Phi_t^{(L)}({\bf r})\right]\right\}\nonumber \\
&+& E[\Phi_h^{(1)}+\Phi_h^{(2)},\Phi_t^{(1)}+\Phi_t^{(2)}]
-{kT\over v_h}\int d{\bf r}\ \Xi({\bf r})
(\Phi_h^{(1)}({\bf r})+\gamma_t\Phi_t^{(1)}({\bf r})+\Phi_h^{(2)}({\bf
r})+\gamma_t\Phi_t^{(2)}({\bf r})-1),
\end{eqnarray}
and ${\cal Q}_L$ is the partition functions of a
single lipid, of type L, in the external fields $W_h^{(L)}$ and $W_t^{(L)}$;
\begin{equation}
{\cal Q}_L=\int\tilde{\cal D}{\bf r}_L\exp\left\{-W_h^{(L)}({\bf
r}_L(\alpha_L/2))-
\int_0^{\alpha_L}ds\ W_t^{(L)}({\bf r}_L(s))\right\}, \qquad L=1,2.
\end{equation}
To this point, no approximations in the evaluation of the partition
function of the model  have been made, but 
it has been put in a form in which the self-consistent field (SCF)
approximation appears naturally. The need for an approximation arises
from the fact that the functional integrals in Eq. \ref{pf2} 
over $W_h^{(L)}$ and $W_t^{(L)}$ cannot be carried out. The SCF consists in
replacing the exact free energy, $-kT\ln{\cal Z}$ by the extremum of 
$\Omega$. We denote the values of the 
$\Phi_h^{(L)}$, $W_h^{(L)}$, $\Phi_t^{(L)}$, $W_t^{(L)}$ and $\Xi$
which extremize $\Omega$ 
by lower case letters. They are obtained from the following 
set of self consistent equations:
\begin{eqnarray}
\label{head}
\phi_h^{(L)}({\bf r})&=&-z_L{\delta{\cal Q}_L\over\delta w_h^{(L)}({\bf
r})},\qquad L=1,2,\\
\label{tail}
\phi_t^{(L)}({\bf r})&=&-z_L{\delta{\cal Q}_L\over\delta w_t^{(L)}({\bf
r})},\qquad L=1,2,\\
w_h^{(L)}({\bf r})&=&2\chi N\sum_{L'}\phi_t^{(L')}({\bf r})-\xi({\bf
r}),\qquad L=1,2,\\
w_t^{(L)}({\bf r})&=&2\chi N\sum_{L'}\phi_h^{(L')}({\bf r})-
\gamma_t\xi({\bf r}),\qquad L=1,2\\
\label{incomp}
1&=&\sum_{L}\phi_h^{(L)}({\bf r})+\gamma_t\sum_L\phi_t^{(L)}({\bf r}).
\end{eqnarray}
Note that $w^{(1)}_h=w^{(2)}_h$ and $w^{(1)}_t=w^{(2)}_t,$ 
so that henceforth we shall drop the
superscripts on the fields $w_h$ and $w_t$. 
It is convenient, further, to introduce the total
headgroup and total tail densities
\begin{eqnarray}
\phi_h({\bf r})&=&\phi_h^{(1)}({\bf r})+\phi_h^{(2)}({\bf r})\\
\phi_t({\bf r})&=&\phi_t^{(1)}({\bf r})+\phi_t^{(2)}({\bf r}),
\end{eqnarray}
and to measure chemical potentails relative to that of lipid 1. This has
the effect that $z_1=1$, and $z_2=z$, the chemical potential of lipid 2
relative to that of 1. 
The free energy in this approximation, 
$\Omega_{scf}$ is
\begin{eqnarray}
-\Omega_{scf}&=&{kT\over v_h}\left[\sum_{L=1}^2z_L{\cal Q}_L[w_h,w_t]
+\int d{\bf r}[w_h({\bf r})\phi_h({\bf r})+w_t({\bf r})\phi_t({\bf r})]\right]
-E[\phi_h,\phi_t],\\
&=&{kT\over v_h}\sum_{L=1}^2z_L{\cal Q}_L[w_h,w_t]+E[\phi_h,\phi_t],\\
\label{omega}
             &=&kT(n_1+n_2)+E[\phi_h,\phi_t],
\end{eqnarray}
with $E$ given by Eq.\ref{energy}.
There remains only the calculation of the partition function of the
single lipid $l$ of type $L$ in the external fields $w_h$ and $w_t$. One
defines the end-segment distribution function
\begin{equation}
q^{(L)}({\bf r},s)=\int{\cal D}{\bf r}_l(s)\delta({\bf r}-{\bf r}_l(s))
\exp\left\{-\int_0^s dt\left(\left[{1\over
8R_g^2}|{d{\bf r}_l(t)\over dt}|^2\right]
+w_h({\bf r}_l(t))\delta(t-\alpha_L/2)+w_t({\bf r}_l(t))\right)\right\},
\end{equation}
which satisfies the equation
\begin{equation}
\label{endpoint}
{\partial q^{(L)}({\bf r},s) \over\partial
s}=2R_g^2\nabla^2q^{(L)}({\bf r},s)-
[w_h({\bf r})\delta(s-\alpha_L/2)+w_t({\bf r})]q^{(L)}({\bf r},s),
\end{equation}
with initial condition
\begin{equation}
q^{(L)}({\bf r},0)=1.
\end{equation}
The partition functions of the two types of lipid are, then,
\begin{equation}
{\cal Q}_L=\int d{\bf r}\ q^{(L)}({\bf r},\alpha_L).
\end{equation}
It then follows from Eqs. \ref{head} and \ref{tail} that
\begin{eqnarray}
\label{head2}
\phi_h({\bf r})&=&\exp[-w_h({\bf r})]\sum_{L=1}^2z_L
q^{(L)}({\bf r},\frac{\alpha_L}{2}-)q^{(L)}({\bf r},\frac{\alpha_L}{2}-),\\
\phi_t({\bf r})&=&\sum_{L=1}^2z_L\int_0^{\alpha_L}ds\ q^{(L)}({\bf r},s)
q^{(L)}({\bf r},\alpha_L-s),\nonumber\\
\label{tail2}
&=&2\sum_{L=1}^2z_L\int_0^{\alpha_L/2}ds\ q^{(L)}({\bf r},s)q^{(L)}
({\bf r},\alpha_L-s),
\end{eqnarray}
where $z_1=1$ and $z_2 = z$. 
The self-consistent equations \ref{head} to \ref{incomp}
can now be solved in real space. For the periodic phases in which we are
interested, such as the lamellar, inverted hexagonal, gyroid, etc., it
is more convenient to do so in Fourier space.
\section{Theory: Fourier Space}
Because the densities, fields, and end point distribution function
depend only on a single coordinate, they reflect the space group
symmetry of the  ordered phase they describe. To make that symmetry
manifest, we expand all functions of position in a complete,
orthonormal, set of
functions $f_i({\bf r}), i=1,2,3...,$ which have the desired space group
symmetry\cite{ms}; {\em e.g.}
\begin{eqnarray}
\phi_h({\bf r})&=&\sum_i\phi_{h,i}f_i({\bf r}),\\
\delta_{i,j}&=&{1\over V}\int d{\bf r}f_i({\bf r})f_j({\bf r}),
\end{eqnarray}
where $V$ is the volume of the system.
Furthermore we choose the $f_i({\bf r})$ to be eigenfunctions of the
Laplacian
\begin{equation}
\nabla^2f_i({\bf r})=-{\lambda_i\over D^2}f_i({\bf r}),
\end{equation} where $D$ is a length scale for the phase.
We set $f_1=1$. 
Expressions for the unnormalized basis functions for all space-group
symmetries can be found in X-ray tables\cite{henry}.
The self-consistent equations in Fourier space become
\begin{eqnarray}
\label{scf1}
w_{h,i}&=&2\chi N\phi_{t,i}-\xi_i,\\
w_{t,i}&=&2\chi N\phi_{h,i}-\gamma_t\xi_i,\\
\delta_{i,1}&=&\phi_{h,i}+\gamma_t\phi_{t,i}.
\end{eqnarray}

To obtain the partition functions and 
densities, we proceed as follows. For any function 
$G({\bf r})$, we can define a symmetric matrix 
\begin{equation}
(G)_{ij}\equiv{1\over V}\int f_i({\bf r})G({\bf r})f_j({\bf r})d{\bf r}
\end{equation}
Note that $(G)_{1i}=(G)_{i1}=G_i$, the coefficient of $f_i({\bf r})$ in
the expansion of $G({\bf r})$. Matrices corresponding to functions of
$G({\bf r})$, such as 
\begin{equation}
\left(e^G\right)_{ij}\equiv{1\over V}
\int f_i({\bf r})e^{G({\bf r})}f_j({\bf r})d{\bf r},
\end{equation}
are evaluated by making an orthogonal transformation which diagonalizes 
$(G)_{ij}$.
The densities are obtained from the end point distribution
function. 
After Fourier transforming the diffusion equation,
Eq. \ref{endpoint}, one readily obtains the solution
\begin{eqnarray}
q^{(L)}_i(s)&=&\left(e^{-A s}\right)_{i,1},\qquad {\rm if}\ 
s<\alpha_L/2\nonumber \\
&=&\sum_j\left(e^{-w_{h}}\right)_{ij}\left(e^{-\alpha_LA/2}\right)_{j,1}
,\qquad s=\alpha_L/2 \nonumber \\
&=&\sum_{j,k}\left(e^{-A(s-\alpha_L/2)}\right)_{i,j}
\left(e^{-w_{h}}\right)_{j,k}\left(e^{-\alpha_LA/2}\right)_{k,1}
, \qquad s>\alpha_L/2,
\end{eqnarray}
where the elements of the matrix $A$ are given by
\begin{equation}
A_{i,j}={2R_g^2\over D^2}\lambda_i\delta_{ij}+(w_t)_{ij}.
\end{equation}
From this, the Fourier amplitudes of the densities follow from
eqs. \ref{head2} and \ref{tail2};
\begin{eqnarray}
\phi_{h;i}&=&\sum_{L=1}^2z_L\sum_{jkl}\left(e^{-w_h}\right)_{ij}
\Gamma_{jkl}q_k^{(L)}({\alpha_L\over2}-)q_l^{(L)}({\alpha_L\over2}-),\\
\label{scf2}
\phi_{t;i}&=&2\sum_{L=1}^2z_L\int_0^{\alpha_L/2} ds\sum_{jk}\Gamma_{ijk}q_j^{(L)}(s)
q_k^{(L)}(\alpha_L-s),\\
\end{eqnarray}
with
\begin{equation}
\Gamma_{ijk}\equiv{1\over V}\int f_i({\bf r})f_j({\bf r})f_k({\bf r}).
\end{equation}
The grand potential within the self-consistent field approximation, 
Eq \ref{omega}, becomes
\begin{equation}
-\Omega_{scf}={kTV\over v_h}\left[\phi_{h,1}+2\chi N\sum_i
\phi_{h,i}\phi_{t,i}\right]
\end{equation} 
This free energy still depends on $D$, the length scale of the periodic
phase, which must be determined by minimization of the free
energy. After this is done, we compare the free energies obtained for
phases of different space group symmetries, and thus obtain the phase
diagram of the system. 

The infinite set of self-consistent equations \ref{scf1} to \ref{scf2}
must be truncated to be solved numerically, and  we have employed up to 50
basis functions in our calculations for the lamellar, hexagonal, and
b.c.c. phases, and 100 for the gyroid phases.

\section{Results}
\subsection{Phase Diagrams}
We begin by presenting, in Fig. 1, our results for the phase diagram of 
a single lipid with
tails each of $N$ units as a function of its architecture. We have
defined an effective temperature, $T^*_N\equiv1/2\chi N$. Note that
$T^*_N$ is length-dependent, i.e., the value of $T^*_N$ differs for lipids
of different length $N$ even at the same physical temperature $T$.
In our previous work\cite{li}, we found that for the system of
dioleoylphosphatidylethanolamine a $T^*_N$ of 0.06 corresponded to a
physical temperature of approximately 20$^{\circ}$C.
The architecture is characterized by the  volume of the
head group relative to that of the entire lipid, 
$f \equiv v_h/(v_h+2Nv_t)$. We have examined the common lipid
phases, including the bi-continuous double-diamond phase (Pn3m),
for stability, and found in addition to the lamellar phase 
only the normal and inverted versions of
the body-centered cubic
($Im3m$), hexagonal, and gyroid
$Ia3d$, phases to be stable. One sees that the diagram is quite
reasonable and illustrates quantitatively the qualitative ideas of
structure being driven by packing considerations\cite{jacob}.

We now turn to a mixture of two lipids. We have chosen $N_1 = N$ and 
$N_2 = 1.5N$ (i.e., $\alpha_1=1$, and $\alpha_2=1.5$) 
so that the extended length of the tails of lipid 2 are 
1.5 times those of lipid 1. Further we have taken $\gamma_t\equiv
2N_1v_t/v_h=2.5$. With these parameters, the volume of the head group 
relative to
that of the entire lipid is, for lipid 1, 
$f_1\equiv 1/(1+\alpha_1\gamma_t)=0.2857$, while that
of lipid 2 is $f_2\equiv1/(1+\alpha_2\gamma_t)=0.2105$. For comparison,
the relative head group volume of dioleoylphosphatidylethanolamine
calculated from volumes given in the literature\cite{randandfuller} is
$f=0.254$. From the phase diagram of Fig. 1, one sees that lipid 1 forms
a lamellar phase, while lipid 2 forms an inverted hexagonal, $H_{II}$,
phase. 

The phase diagram of the lipid mixture is shown in the solid lines of
Fig. 2 as a function
of the volume
fraction of lipid 1, $\Theta\equiv\phi_h^{(1)}+\gamma_t\phi_t^{(1)}$ and 
the reduced temperature $T^*\equiv 1/2\chi N_1$. Here $N_1$ is a
constant, 
the length of the tails of lipid 1, so that the definition of the
temperature $T^*$ is independent of the composition of the mixture as it
should be.
Small regions of
body-centered cubic phase near the transition to the disordered phase
have been ignored in the phase diagram. We are unaware of experimental
phase diagrams from anhydrous mixtures of lamellar- and
hexagonal-forming lipids with the same headgroup,
as we have calculated here. However phase diagrams have been obtained for
mixtures of  lamellar-forming phosphatidylcholine 
and hexagonal-forming phosphatidylethanolamine. Those obtained for 
anhydrous mixtures of
dilinoleoylphosphatidylethanolamine and
palmitoyloleoylphosphatidylcholine\cite{boni}, 
and for mixtures of dioleoylphosphatidylcholine and
dioleoylphosphatidylethanolamine and 10\% water by  
weight\cite{per} are quite similar to Fig. 2 as they each 
show a significant region of
cubic phase between the inverted hexagonal phase, which dominates at low
concentrations of the lamellar-forming lipid, and the lamellar phase,
which dominates at high concentration. 
We also observe
that about 20\% of one lipid added to the other is sufficient to bring
about a change of phase, which is in accord with experiment\cite{tilcock}. 
Lastly, if the temperature is not too low, a decrease of
the volume fraction of the lamellar-forming lipid causes the gyroid
phase to be stable at increasing temperatures. Therefore the addition of
longer non-lamellar lipids to the mixture stabilizes the non-lamellar phase,
as in the experiment
of Tate and Gruner\cite{tate}.

We now show that there is a simple ``effective single lipid"
approximation by which one can obtain rather well the phase diagram 
of the mixture, Fig. 2,
utilizing only the information from the phase diagram of the
single lipid, Fig. 1. For this purpose, we must find a relationship between
the coordinates $(f, T^*_N)$ in Fig.1 and $(\Theta, T^*)$ in Fig. 2. 
Note that whereas the temperature scale $T^*$ is independent of the
composition of the mixture, the appropriate scale of Fig. 1 would vary
from $T^*_{N_2} = T^*/1.5$ when the mixture contained lipid 2 only 
to $T^*_{N_1} = T^*$ at the other extreme. 

To obtain the desired relationship, we first note that from the
definitions of $f$ and $\gamma_t$, it follows that 
$2 N v_t/v_h = (1-f)/f$ and $2 N_1 v_t/v_h = \gamma_t$,  from which one obtains
$T^*_N/T^* = N_1/N = f \gamma_t/(1-f)$. 
Second, as $f$ is defined as the volume fraction of the head group 
for a single lipid, it is natural to assume $f = \Theta f_1 + (1-\Theta) f_2$,
the volume fraction of the head groups, in the mixture. 
Hence, in the ``effective single lipid" approximation 
the coordinates of a point on the phase diagram of the mixture may be 
obtained from those of the single lipid according to
\begin{eqnarray}
\Theta&=&\frac{f-f_2}{f_1-f_2},\\
T^*&=&\frac{1-f}{f\gamma_t}T^*_N.
\end{eqnarray}
The results of this ``single effective lipid" approximation is shown by
the open diamonds in Fig. 2. The approximation is obviously very good.
Thus we can obtain very easily the phase diagram of any
anhydrous mixture of our model lipids from the results for a single
lipid, given in Fig. 1.

\subsection{Distribution of Lipids}
We now turn to the central results of this paper, which is the
distribution of the different lipids in the various phases. It is
convenient at the outset to define two local order parameters, $\psi({\bf
r})$ and $\zeta({\bf r})$:
\begin{eqnarray}
\psi({\bf r})&=&\psi^{(1)}({\bf r})+
\psi^{(2)}({\bf r}),\\ 
\psi^{(L)}({\bf r})&\equiv&
\phi_h^{(L)}({\bf r})-\frac{\phi_t^{(L)}({\bf r})}{\alpha_L},\qquad L=1,2,\\
\zeta({\bf r})&\equiv&\frac{\phi_t^{(1)}({\bf r})}{<\phi_t^{(1)}({\bf r})>}-
\frac{\phi_t^{(2)}({\bf r})}{<\phi_t^{(2)}({\bf r})>}.
\end{eqnarray} 
Each order
parameter, $\psi^{(L)}({\bf r})$ measures the local difference in head 
and tail segments of type $L$ normalized such that the integral of the
order parameter over the unit cell vanishes. Thus $\psi({\bf r})$
measures the local difference of all head and tail segments. It provides
information on the separation  of the lipid heads and tails. On the
other hand, the order
parameter $\zeta({\bf r})$ measures the difference in local fractions of
the tail segments belonging to the two different lipids. The brackets in
its definitions denotes an average over some suitably defined tail
region in which the order parameter $\psi({\bf r})$ takes on negative
values. For clarity, we employ somewhat different definitions of the
tail region for the different phases. The average value of 
$\zeta({\bf r})$ is zero over the defined tail region.

We begin
with the lamellar phase and show in Fig. 3 the density profiles at a
temperature $T^*=0.0454$ and volume fraction of lipid 1, $\Theta=0.6588$, a 
point very close to the phase boundaries between 
the lamellar phase and the inverted gyroid, $G_{II}$, phase. 
In Fig 3(a) we have plotted the volume fractions of the heads of
lipids 1 and 2, $\phi_h^{(1)}$ and  $\phi_h^{(2)}$, and the volume
fractions of the tails, $\gamma_t\phi_t^{(1)}$, and 
$\gamma_t\phi_t^{(2)}$. Although the lamellar forming lipid 1
dominates, one can easily see that the density of lipid 1 tails
decreases near the center of the lamellae by the order of 5 per cent
from its maximum value of about 0.6. 
As the system is incompressible, this implies that
the volume fraction of the tails of the non-lamellar forming lipid
increases by about 8 per cent over the same range.
To illustrate the {\em relative} difference in their
volume fractions, we plot in Fig. 3(b) the order parameter $\zeta(x)$ 
where the tail region is defined here as $\psi(x) \le 0$. Again one
sees that the relative change in volume fractions is less than, but the
order of 10 per cent, with the
longer lipid 2 predominating in the center of the tail region. This
comes about for two reasons. First, the tails of lipid 1 are  shorter
than those of lipid 2, so
one expects there to be less lipid 1 near the center. Second, the
effect of the temperature, which is to shorten the average end to end
distance of the tails, is greater on the tails of lipid 1 than on the
tails of lipid 2 because $T^*_{N_1} > T^*_{N_2}$. We have not
plotted the relative variation of the head group volume fractions
because it is so small, of the order of 0.5 per cent. 
Recall that all the head groups are identical.

We now turn to the inverted hexagonal, $H_{II}$, phase. We consider the
system again at $T^*=0.0454$ but now at volume fraction of lipid 1 of
$\Theta=0.27$ where the $H_{II}$ and gyroid phases are almost in coexistence. 
In Fig. 4, we
plot the order parameter $\psi({\bf r})$. The dark
regions correspond to positive values of the order parameter where the
head groups dominate and the lighter regions where the tails
dominate. The maximum value is 0.719 and the minimum value $-0.286$. Each
gradation represents a change of 10 per cent. To make manifest the
difference in densities of the tails from each lipid, we again look at
the order parameter $\zeta({\bf r})$ which is shown in Fig. 5. Here the
tail region we have averaged over is the locus of points for which the
order parameter $\psi({\bf r})\leq-0.282$. Regions with 
$\psi({\bf r}) > -0.282$ are simply shown in white.
Again the darker regions
correspond to positive values of this order parameter where the tails of
the lamellar forming lipid are relatively more probable to be found,
while the lighter regions denote  negative values where the non-lamellar
forming lipid is more probable. One see that the latter lipid is more
likely to be found in the next-nearest neighbor direction, in the region
which is most difficult for the tails to fill. The maximum value of
$\zeta({\bf r})$ is 0.0352, and the minimum value is $-0.0377$ indicating that
the relative volume fractions vary  by the order of 3 per cent over the
tail region. To make the
variation in densities even clearer, we plot in Fig. 6(a) and 6(b) the 
volume fractions of the tails of the  lipids as measured along the
boundary of the Brillouin zone with the angle $\theta=0$ corresponding
to the direction of the nearest neighbors, and $\theta_0=\pi/6$ corresponding
to the direction of the next nearest neighbors. One sees that the
density of the minority lamellar forming lipid is reduced by about 2 per
cent in
the next-nearest neighbor direction, while that of the majority non-lamellar
forming lipid is is increased by about 0.6 per cent. Again, the
variation with direction  of the density of head groups is negligible.
 
In order to compare these densities changes with the curvature of the
inverted hexagonal phase, we plot in Fig. 6(c) the mean curvature along
the locus of points defined by $\psi({\bf r})=0$, {\em i.e.}, on the
curve on which the difference in volume fractions of all heads and tails
vanishes. The mean curvature as a function of angle is well fit by
\begin{equation}
H(\theta)=H_0+\sum_{n=1}^{5}H_n\cos(6n\theta),
\end{equation}
with $H_0=1.55800$, $H_1=-3.33\times10^{-3}$, $H_2=3.2\times 10^{-4}$,
$H_3=-5.3\times10^{-4}$,$H_4=3.59\times10^{-3}$, and
$H_5=-8.2\times10^{-4}$. One see that the variation in the mean
curvature is an order of magnitude smaller than that of the variation of
the tails of 
the minority lipid, and also smaller than that of the majority lipid. 
There is also much more structure in the mean curvature. It is not
surprising that this structure near the region where the head groups and
tails meet is washed out in the region of the other end of the tails.

We now consider the inverted gyroid phase. We show in Fig. 7 and 8 two
views of this phase at $T^*=0.045$ and $\Theta=0.4921$. The phase 
consists of two sublattices of tubes filled
for the most part with head groups. The sublattices are related by
mirror reflection. We have chosen to plot the surfaces defined by
the order parameter $\psi({\bf r})=0.5$ rather than $\psi({\bf r})=0$ 
for in the latter case the tubes are much more difficult to recognize. Fig. 7
shows a view along the [001] direction while Fig. 8 is viewed from point
(0.9,-2.4,2). 
These figures define the coordinate system which we use and show a
cubic cell of side unity. Were the tubes shrunk to lines, three such
lines would meet at nodes, and there would be 16 such nodes in the unit
cell shown.

We wish to show what the relative densities of the two lipids are at
the positions which are the easiest for the tails to fill, and those
which are the most difficult for the tails to fill. These correspond to
the points midway between nearest neighbors and next-nearest neighbors
in the $H_{II}$ phase. These positions are not obvious in the gyroid
phase, but can be calculated. An example of a longest distance 
is shown in Fig. 8 by a solid line. The coordinates of the point shown,
which is 
furtherest from any tube, is $(1/2,3/16,5/8)$ while the points on the
tube centers closest to it are $(15/32,7/32,7/8)$ and $(17/32,7/32,3/8)$.
Thus this longest distance is $d_{long}=(66)^{1/2}/32$. The shortest distances
link two three-fold coordinated sites on the different sublattice, and
one such shortest distance is shown in Fig. 8 by a dotted line. Two
particular points on the tube centers connected by this distance and
which are shown in the
figure have the coordinates $(5/8,3/8,7/8)$ and $(7/8,1/8,5/8)$. The
distance between these points is twice the shortest distance between a
point midway between tubes and the tube centers themselves, and is therefore
$d_{short}=3^{1/2}/8$.  
 Interestingly, the ratio of longest to shortest
distances in the gyroid phase
$d_{long}/d_{short}=(33/8)^{1/2}/(3)^{1/2}$ 
is only about 2 per cent larger than the corresponding 
ratio in the hexagonal and b.c.c. phases which is
$2/(3)^{1/2}$ in both cases. At the temperature and composition
selected, we find that at the point nearest to neighboring tubes and
which requires the least stretching of the lipid tails, 
the volume fraction of the tails of
the lamellar forming lipid 1 is $\gamma_t\phi_t^{(1)}=0.448$ and that of
the non-lamellar forming lipid 2 is $\gamma_t\phi_t^{(2)}=0.537$. 
At the position requiring the most stretching of the tails, we find 
the volume fraction of the tails of lipid 1 
has decreased to 0.431, while that of the longer lipid 2 has increased 
to 0.561. 

In Fig. 9 the order parameter $\zeta({\bf r})$, which shows the
relative difference
between the volume fractions of the tails of lipid 1 and lipid 2, is
shown
in a  cut through the gyroid taken in the [$1{\bar 1} 0$] direction and which
passes through two points, $(-1/8,-1/8,-1/8)$ and $(1/8,1/8,1/8)$, which 
require the least stretch. Only the portion
of the tail region
defined by $\psi({\bf r})\leq-0.2$ is shown with any gray scale variation. 
The solid line connects the tube centers which lie closest to one
another. The point midway along this line is that most easily reached by
the lipid tails. The fact that the region around this point 
is dark indicates that the shorter
lipid 1 is relatively more probable here. The maximum (black) and
minimum values (white) of $\zeta({\bf r})$  are $9.26\times 10^{-2}$ and
$-6.64\times 10^{-2}$ indicating a variation of relative volume fraction 
in the tail region shown which is on the order of 8 per cent. 
Figure 10 shows a cut in the [801]
direction which passes through the point which is furtherest from the
center of any tube. Again the region shown in gray scale variation is
the tail region defined by $\psi({\bf r})\leq -0.2$. 
The maximum and minimum values of $\zeta({\bf r})$ are $10.35\times
10^{-2}$ and $-6.99\times 10^{-2}$ indicating a similar variation in
volume fraction. The point at the center of the bent dark line is
furtherest from the center of any tube. The black line ends on the 
centers of the two tubes closest to it. The lightest part of the
generally dark tail region is very close to the point furtherest from
any tube center and
indicates that the relative probability of finding the tails of the
non-lamellar forming lipid is large there. That the lightest part does
not correspond exactly to the point furtherest from the tube center is
probably due to the fact that this point is not necessarily the
furtherest from the head-tail interface, and thus
not necessarily the most difficult for the tails to reach, although
it is probably quite close to being so.

In Fig. 11 we show the surface  defined by the order parameter
$\psi({\bf r})=-0.2$ in the region of the three-fold connectors. On that
surface the order parameter $\zeta({\bf r})$ is plotted. The largest
value, which is black and at the center of this region, is
$3.25\times10^{-2}$ showing that the lamellar-forming lipids are more
probable there. The smallest  value, shown in the lightest gray, is
$-1.82\times10^{-2}$ indicating a few per cent variation in this region.
For comparison, the mean curvature on the surface defined by $\psi({\bf
r})=0$ is plotted in Fig. 12. The smallest value of the mean curvature, $0.20$,
occurs at the center, while the largest values, $0.52$, occur away from
it. As in the inverted hexagonal phase, one sees that the preferential
location of the lamellar-forming lipids is at low curvature
sites. Further there is more structure in the plot of the curvature on a
surface near where the heads and tails meet than in the plot of the 
distribution of tail volume fractions deep within the tail region. 
That the region near the center of these three-fold connectors is
characterized by small mean curvature is analogous to the results
obtained by Matsen\cite{mat} for the gyroid phase of block copolymers.
We used two hundred basis functions for this plot.

In sum, we have employed a lipid model which has given excellent results
previously\cite{li} for the phase diagram of 
dioleoylphosphatidyethanolamine in order to examine the distribution of
tails in a mixture of  lamellar forming and non-lamellar forming lipids.
The two lipids are characterized by the same head group, but tails of
different length. We have shown, both in the inverted hexagonal and
gyroid phases, that the tails of the non-lamellar
forming lipid are found preferentially in regions of the unit cell which
are difficult to fill, and those of the lamellar forming lipid are found
in the regions most easily filled. The variation in volume fraction is of the
order of 1 to 10 per cent. The difference in lipid tail density is also
correlated with the curvature of the surface which, loosely, separates
the head groups from the tails although the former shows much less
structure than the latter. It will be most interesting to apply this
model to stalk-like structures thought to be of importance in membrane
fusion to determine by what amount the presence of non-lamellar forming
lipids can lower the free energy barrier to their formation\cite{siegel1}.
This work was supported in part by the National Science Foundation under
grant number DMR9876864.

\newpage

\begin{description}
\item[Figure 1.] Phase diagram of an anhydrous system of lipids as a
function of the temperature $T^*_N$ and the volume of the head group
relative to that of the entire lipid, $f\equiv
v_n/(v_n+2Nv_t)=1/(1+\gamma_t)$. In addition to
the disordered ($D$)  and lamellar $L_{\alpha}$ phases, there are
body-centered cubic ($bcc$), hexagonal ($H$), and gyroid $G$ phases. The
subscripts $I$ and $II$ denote normal and inverted phases respectively.
The dashed lines indicate extrapolated boundaries.
\item[Figure 2.] Phase diagram of a mixture of two lipids with the same
head group, but one with tails 1.5 times the length of the other. Lipid
1, a lamellar-forming lipid, is characterized by $1/(1+\gamma_t)=0.2857$
while the non-lamellar forming lipid 2, 
with the longer tails, is characterized by
$1/(1+1.5\gamma_t)=0.2105$. $T^*=1/2\chi N_1$ is a measure of the 
temperature, and $\Theta$ is the volume fraction of lipid 1. 
Solid lines result from the full
self-consistent field calculation of the mixture, while open diamonds
result from the single effective lipid approximation. Very small regions
of b.c.c phases have been ignored.
\item[Figure 3.](a) Volume fractions of the head groups $\phi_h^{(1)}$ and
$\phi_h^{(2)}$  of lipids 1 and 2 and of the tail groups
$\gamma_t\phi_h^{(1)}$ and $\gamma_t\phi_h^{(2)}$ in the lamellar phase
at $T^*=0.045$ and volume fraction of lipid 1 $\Theta=0.6588$.

(b) The order parameter $\zeta(x)\equiv\phi_t^{(1)}(x)/
<\phi_t^{(1)}(x)>-\phi_t^{(2)}(x)/<\phi_t^{(2)}(x)>$, where the averages are
taken over the region $\psi(x) \le 0$.
\item[Figure 4.] The order parameter $\psi({\bf r})
=\phi_h^{(1)}({\bf r})-\phi_t^{(1)}({\bf r})
 +\phi_h^{(2)}({\bf r})-(1/\alpha_2)\phi_t^{(2)}({\bf r})$ in the
$H_{II}$ phase at $T^*=0.045$ and $\Theta=0.27$. 
The maximum value is 0.719 and the minimum value $-0.286$. Each
gradation represents a change of 10 per cent. 
\item[Figure 5.] The order parameter $\zeta({\bf r})$ in the hexagonal
phase of Fig. 4. The averages are taken over the region
$\psi({\bf r})\leq-0.282$. Regions with $\psi({\bf r})>-0.282$ are 
shown in white. Light parts of the tail region indicate an excess of
non-lamellar forming lipids, while darker areas represent excess of
lamellar-forming lipids.
\item[Figure 6.] (a) Volume fraction of non-lamellar forming lipid 2
evaluated on the Brillouin zone edge from nearest-neighbor direction,
$\theta=0$,
 to
next-nearest-neighbor direction, $\theta=\pi/6.$ 

(b) Volume fraction of lamellar forming lipid 1
evaluated on the Brillouin zone edge.

(c) Mean curvature of the surface defined by $\psi({\bf r})=0$.
\item[Figure 7.] A view of the gyroid taken along the $[001]$
direction. The surface is defined by the value of the order parameter
$\psi({\bf r})=0.5$ which is well within the region in which the
head groups dominate.
\item[Figure 8.] A second view of the gyroid surface, defined as in 
Fig. 7. The longest distance between a point in the tail
region and the axis of any tube is shown by two intersecting solid
lines. The shortest distance between the axes of two tubes 
is shown by a dotted line. Such
a line connects nearby regions of three-fold symmetry on the two
different sublattices. 
\item[Figure 9.]  Order parameter $\zeta({\bf r})$ shown in a cut which
passes through the point requiring the least stretch of the lipid
tails. The points closest together are shown connected by a solid line. 
The volume fractions of the tails are averaged over the region defined by
$\psi({\bf r})\leq-0.2$. The region in which $\psi({\bf r})>-0.2$ is
shown in white. The maximum value of $\zeta({\bf r})$, in black, 
is $9.26\times10^{-2}$ and the minimum value is $-6.64\times10^{-2}$.
\item[Figure 10.]
Order parameter $\zeta({\bf r})$ shown in a cut which
passes through the point furtherest from the axis of any tube. 
The volume fractions of the tails are averaged over the region defined by
$\psi({\bf r})\leq-0.2$.The maximum value of $\zeta$ is $10.35\times10^{-2}$ and
the minimum is $-6.99\times 10^{-2}$.
\item[Figure 11.] The order parameter $\zeta({\bf r})$ is plotted on the 
surface  defined by the order parameter
$\psi({\bf r})=-0.2$ in the region of the three-fold connectors.
The largest
value, which is black and at the center of this region, is
$3.25\times10^{-2}$ showing that the lamellar-forming lipids are more
probable there. The smallest  value, shown in the lightest gray, is
$-1.82\times10^{-2}$ indicating a few per cent variation in this region.
\item[Figure 12.] The mean curvature on the surface defined by $\psi({\bf
r})=0$ is shown. The smallest value of the mean curvature, $0.20$,
occurs at the center, while the largest values, $0.52$, occur away from
it.
\end{description}
\end{document}